\documentclass[a4paper, 11pt]{article}

\usepackage[arxiv]{optional}
\usepackage[english]{babel}
\usepackage[T1]{fontenc}
\usepackage{amsmath}
\opt{arxiv}{%
\usepackage{amsthm}
}%
\usepackage{amsfonts}
\usepackage{amssymb}
\usepackage[final]{graphicx}
\usepackage{paralist}
\opt{draft}{%
\usepackage{refcheck}
}

\opt{arxiv}{%
\theoremstyle{plain}
\newtheorem{thm}{Theorem}[section]
\newtheorem{prop}[thm]{Proposition}
\newtheorem*{prop*}{Proposition}
\newtheorem{lem}[thm]{Lemma}
\newtheorem*{lem*}{Lemma}
\newtheorem{cor}[thm]{Corollary}
\newtheorem*{cor*}{Corollary}

\newtheorem*{example*}{Example}
\newtheorem{conject}[thm]{Conjecture}
\newtheorem*{conject*}{Conjecture}

\theoremstyle{definition}
\newtheorem{defn}[thm]{Definition}
\newtheorem*{defn*}{Definition}
\newtheorem{rem}[thm]{Remark}
\newtheorem*{rem*}{Remark}
}%
\opt{springer}{%

}%

\newcommand{\eps}{\varepsilon}

\opt{springer}{%

}
\opt{arxiv}{%

}

\newcommand{\abs}[1]{\vert #1 \vert}

\newcommand{\C}{\ensuremath{\mathbb{C}}}

\newcommand{\conj}[1]{\overline{#1}}

\newcommand{\jac}{D}
\newcommand{\derz}{\partial_{z}}
\newcommand{\derc}{\partial_{\bar{z}}}

\title{Creating images by adding masses to gravitational point lenses}

\author{Olivier S\`{e}te, Robert Luce, J\"org Liesen\footnote{TU
Berlin, Institute of Mathematics, MA 4-5, Stra{\ss}e des 17. Juni 136,
10623 Berlin, Germany (\texttt{\{sete,luce,liesen\}@math.tu-berlin.de})}}

\begin{document}
\maketitle

\begin{abstract}
A well-studied maximal gravitational point lens construction of S.~H.~Rhie
produces $5n$ images of a light source using $n+1$ deflector masses.
The construction arises from a circular, symmetric deflector
configuration on $n$ masses (producing only $3n+1$ images) by adding
a tiny mass in the center of the other mass positions (and reducing
all the other masses a little bit).

In a recent paper we studied this ``image creating effect'' from a
purely mathematical point of view (S\`ete, Luce \& Liesen, Comput.\@
Methods Funct.\@ Theory 15(1):9-35, 2015).  Here we discuss
a few consequences of our findings
for gravitational microlensing models.  We present a complete
characterization of the effect of adding small masses to these point
lens models, with respect to the number of images.
In particular, we give several examples of maximal lensing models that
are different from Rhie's construction and that do not share its
highly symmetric appearance.  We give generally applicable conditions
that allow the construction of maximal point lenses on $n+1$ masses
from maximal lenses on $n$ masses.

\end{abstract}

\section{Introduction}

We consider the phenomenon of multiple lensed images in the framework
of gravitational microlensing.  Specifically, given $n \ge 2$ point masses
$m_j > 0$ at positions $z_j \in \C$ in the complexified lens plane, we
consider the lensing map $\eta : L \rightarrow S$ from the lens plane
$L = \C \backslash \{ z_1, \ldots, z_n \}$ to the light source plane
$S = \C$,
\begin{equation}
\label{eqn:lens_map}
    \eta(z) = z - \gamma \conj{z}
        - \sum_{j=1}^n \tfrac{m_j}{\conj{z} - \conj{z}_j},
\end{equation}
where $\gamma \in \C$ is the (constant) external shear.  This lens
model can be seen as a generalization of the Chang-Refsdal lens to $n$
point massses; see~\cite{AnEvans2006}.  Given a light source position
(projected on the lens plane) $\zeta \in \C$, the (projected) images
of the light source are exactly the solutions of the equation $\eta(z)
= \zeta$. See~\cite{Wambsganss1998} for a general introduction to
gravitational lensing and~\cite{Straumann1997} for gravitational
lensing in terms of complex variables; see
also~\cite{PettersLevineWambsganss2001,KhavinsonNeumann2008}.

The important question of the maximal number of images that can be
produced by a gravitational lens on $n$  point masses modeled
by~\eqref{eqn:lens_map} was answered in 2006 by Khavinson \&
Neumann~\cite{KhavinsonNeumann2006}.  Their result is as follows.
\begin{thm}
The maximal number of images that can be produced by the lensing map
$\eta$ in~\eqref{eqn:lens_map} is $5n-5$ if $\gamma = 0$ and $5n$ if
$\gamma \neq 0$.
\end{thm}

In the case of \emph{nonzero} shear, the bound of $5n$ images can be
improved slightly.  As shown by An \& Evans~\cite{AnEvans2006}, the
maximal number of images in that case is $5n - 1$.  The ``missing
image'' accounts for a solution to the lens equation at the point
infinity in the extended complex plane; see
also~\cite{LuceSeteLiesen2014b}.

A particular class of point lenses that realizes the maximal number of
images has been devised by Rhie~\cite{Rhie2003}.  Her construction
(and the variant discussed in~\cite{BayerDyer2007,BayerDyer2009}) has
been recently studied in great detail~\cite{LuceSeteLiesen2014a}. We
will very briefly recall the construction with a small \emph{additive}
mass (in contrast to her original construction in~\cite{Rhie2003}).

Consider the lens on $n$ equal point masses $m_j = 1/n$ located at the
vertices of a regular polygon of a certain radius $r$, i.e., $z_j = r
e^{i \frac{2j\pi}{n}}$, and without external shear ($\gamma=0$).  This
yields the lensing map
\begin{equation}
\label{eqn:MWP_lens}
    \eta(z) = z - \tfrac{\conj{z}^{n-1}}{\conj{z}^n - r^n}.
\end{equation}
For a light source located at the origin of the lens plane, i.e.,
$\zeta = 0$, it is known that this lens produces $3n+1$
images~\cite{MaoPettersWitt1999}.
In order to arrive at a maximal lens, a tiny mass $\eps$ is added at the
image position $z=0$, i.e., we define
\begin{equation}
\label{eqn:BD_lens}
    \eta_\eps(z) = \eta(z) - \tfrac{\eps}{\conj{z}}.
\end{equation}
If $\eps > 0$ is sufficiently small, this ``perturbation'' of the lens
induces $2n$ ``new'' images on two circles around the
origin~\cite{LuceSeteLiesen2014a} (and the previous image at $z=0$
vanishes). So the lens on $n+1$ point masses modeld by $\eta_\eps$
produces $5n$ images, and thus is a maximal lens.

We recently showed (in a purely mathematical context) that this
``image creating effect'' of adding masses is \emph{not specific} to
the particular (symmetric) lens described
by~\eqref{eqn:MWP_lens}~\cite{SeteLuceLiesen2014}.  Our goal here is
to present some implications of the mathematical results
in~\cite{SeteLuceLiesen2014} for gravitational point lens models.

In Section~\ref{sec:classification} we present a general
classification of the image creating effect that is induced by adding
tiny masses to an existing lens.  The results are applicable to point
lens models with or without external shear.
The extremal case of \emph{maximal lensing} is studied in
Section~\ref{sec:maximal_lensing}.  To our knowledge, the
only known maximal point lens models are based on the
lens~\eqref{eqn:MWP_lens} from above.  We will present conditions that
allow the construction of maximal point lenses different from these
lenses.  We give several examples for maximal lenses.

\section{Adding tiny masses to a lens}
\label{sec:classification}

Recall that the solutions to the lens equation $\eta(z) = \zeta$ can be
classified using the sign of the determinant of the Jacobian of
$\eta$ (e.g.~\cite{PettersWerner2010}). In terms of Wirtinger
derivatives, we find for the functional determinant of the lensing map
\begin{equation*}
   \det \jac \eta(z) = \abs{\derz \eta(z)}^2 - \abs{\derc \eta(z)}^2
        = 1 - \abs{R'(z)}^2,
\end{equation*}
where we have abbreviated $R(z) = \conj{\gamma} z + \sum_{j=1}^n \tfrac{m_j}{z
- z_j}$, so that $\eta(z) = z - \conj{R(z)}$.

We will call an image $z^* \in \C$, i.e, a solution to the equation
$\eta(z) = \zeta$, a \emph{sense-preserving} image if $\abs{R'(z^*)}
< 1$, and a \emph{sense-reserving} image if $\abs{R'(z^*)} > 1$.  The 
sense-reversing images correspond to saddle images (i.e., the Jacobian
of $\eta$ is indefinite), whereas sense-preserving images correspond
to minimum or maximum images (where the Jacobian is definite).
Recall that an image is 
called a minimal, saddle or maximal image if it is a (local) minimum, saddle 
point or (local) maximum of the time delay function (induced from the 
lens potential corresponding to $\eta$); see
e.g.~\cite{PettersWerner2010,PettersLevineWambsganss2001}.
The characterization via $\abs{R'}$ then follows from the equality of the 
Jacobian determinant of the lensing map and the determinant of the Hessian of 
the time delay function.

Note that the functional determinant vanishes at an image $z^*$ only
if $\zeta$ lies on a caustic (i.e., infinite magnification), and we will
assume in the following that this is not the case.  Further we will
assume in the following that $\abs{\gamma} \neq 1$.

We will now rephrase Theorems 3.1 and 3.14
of~\cite{SeteLuceLiesen2014} into the setting of gravitational
microlensing.  In short, the following theorem can be summarized as
follows: If a sufficiently small mass is inserted at position
$z_{n+1}$ of the lens plane, there will always appear some ``new''
images nearby $z_{n+1}$, and all the previously existing images will
only alter their positions slightly (except for possibly $z_{n+1}$, if
it is an image position itself).  The number of new images depends on
certain properties of the lensing map at $z_{n+1}$, which we can fully
classify.

\begin{thm}
\label{thm:pert}
Let $\eta_n(z) = z - \gamma \conj{z} - \sum_{j=1}^n \frac{m_j}{\conj{z}
- \conj{z}_j} = z - \conj{R(z)}$ be the lensing map corresponding to
  $n \ge 2$ point masses $m_j > 0$ at positions $z_j \in \C$ with external shear
$\abs{\gamma} \neq 1$.  Denote by $m_{n+1} > 0$ a tiny mass and $z_{n+1} \in
\C$, $z_{n+1} \neq z_j$ for $1 \le j \le n$, a point on the lens plane at
which the mass is added, i.e., consider the lensing map
\begin{equation*}
    \eta_{n+1}(z) = \eta_n(z) - \tfrac{m_{n+1}}{\conj{z} - \conj{z}_{n+1}}.
\end{equation*}
If $m_{n+1}$ is sufficiently small, and if the source $\zeta$ does not
lie on a caustic of $\eta_n$ or $\eta_{n+1}$, then there exists an
open disk $D$ around $z_{n+1}$ such that $D \setminus \{ z_{n+1} \}$
contains no mass point of $\eta_n$ 
and no image of $\zeta$ under $\eta_n$.  Further $\eta_n$ and 
$\eta_{n+1}$ have the same number of images outside $D$, and the
following holds:
\begin{compactenum}

\item If $z_{n+1}$ is not an image of $\zeta$ under $\eta_n$, then
$\eta_{n+1}$ has at least one image of $\zeta$ in $D$.\label{case:no_img}

\item If $z_{n+1}$ is a sense-reversing image of $\zeta$ under $\eta_n$,
then $\eta_{n+1}$ has at least two images of $\zeta$ in $D$.\label{case:sr_img}

\item If $z_{n+1}$ is a sense-preserving image of $\zeta$ under $\eta_n$,
and $\abs{R'(z_{n+1})} \neq 0$, then $\eta_{n+1}$ has at least four images
of $\zeta$ in $D$.\label{case:sp_img}

\item If $z_{n+1}$ is a sense-preserving image of $\zeta$ under $\eta_n$,
$R'(z_{n+1}) = \dotsb = R^{(d-2)}(z_{n+1}) = 0$, and $R^{(d-1)}(z_{n+1})
\neq 0$, then $\eta_{n+1}$ has at least $2d$ images
of $\zeta$ in $D$.\label{case:sp_img_zeroder}

\end{compactenum}
\end{thm}

\begin{figure}[t]
\includegraphics[width=.48\textwidth]{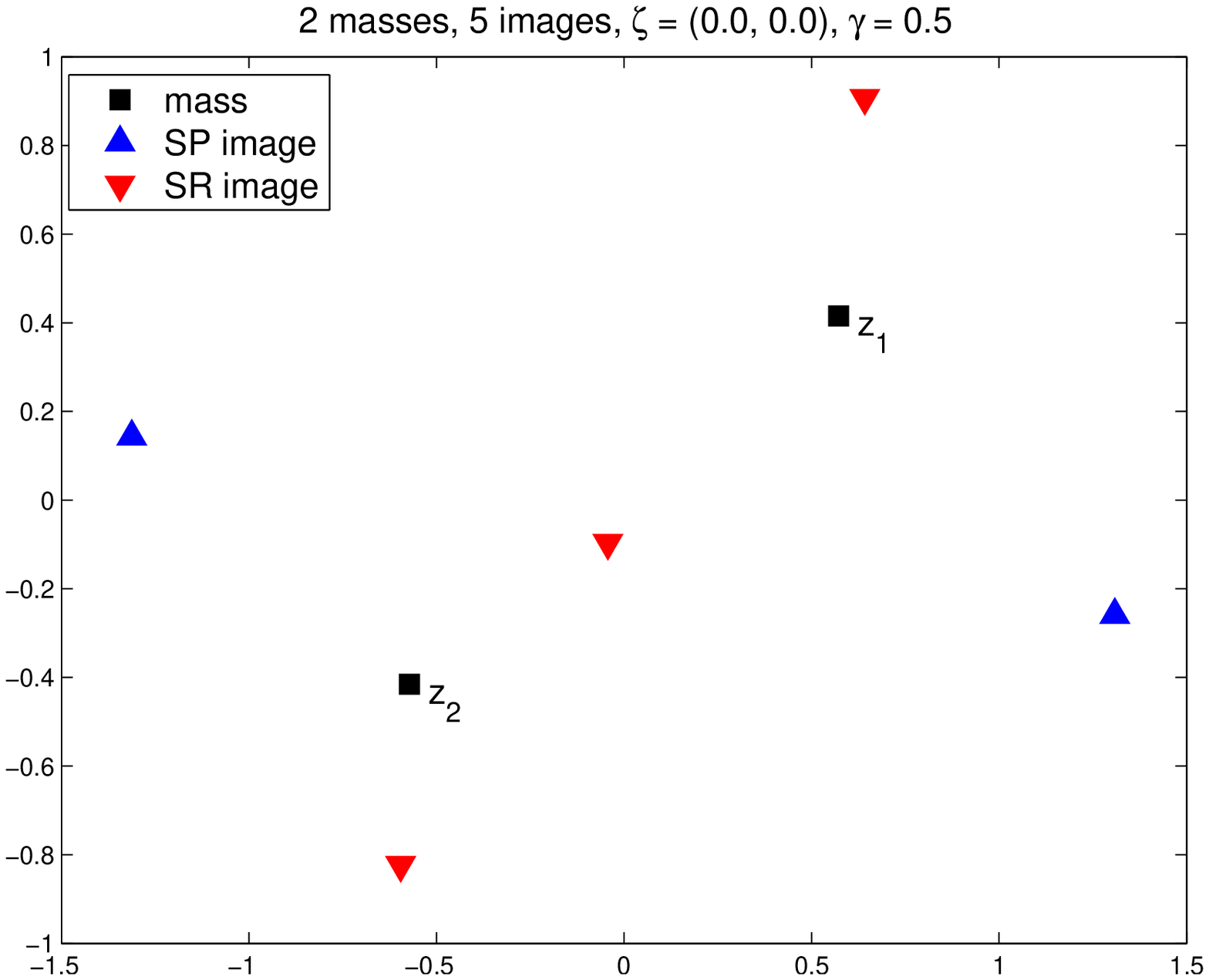}
\hfill
\includegraphics[width=.48\textwidth]{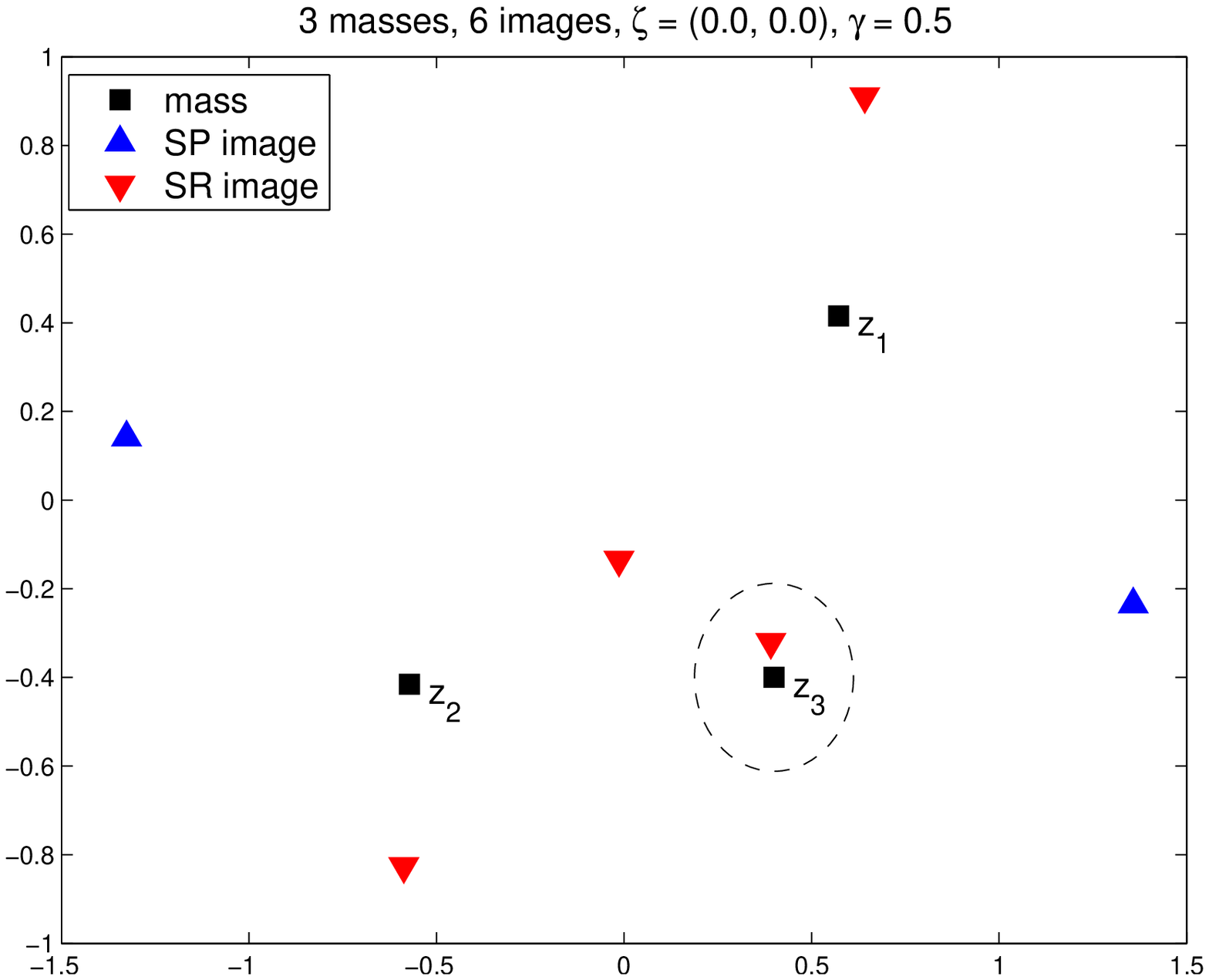}
\newline
\includegraphics[width=.48\textwidth]{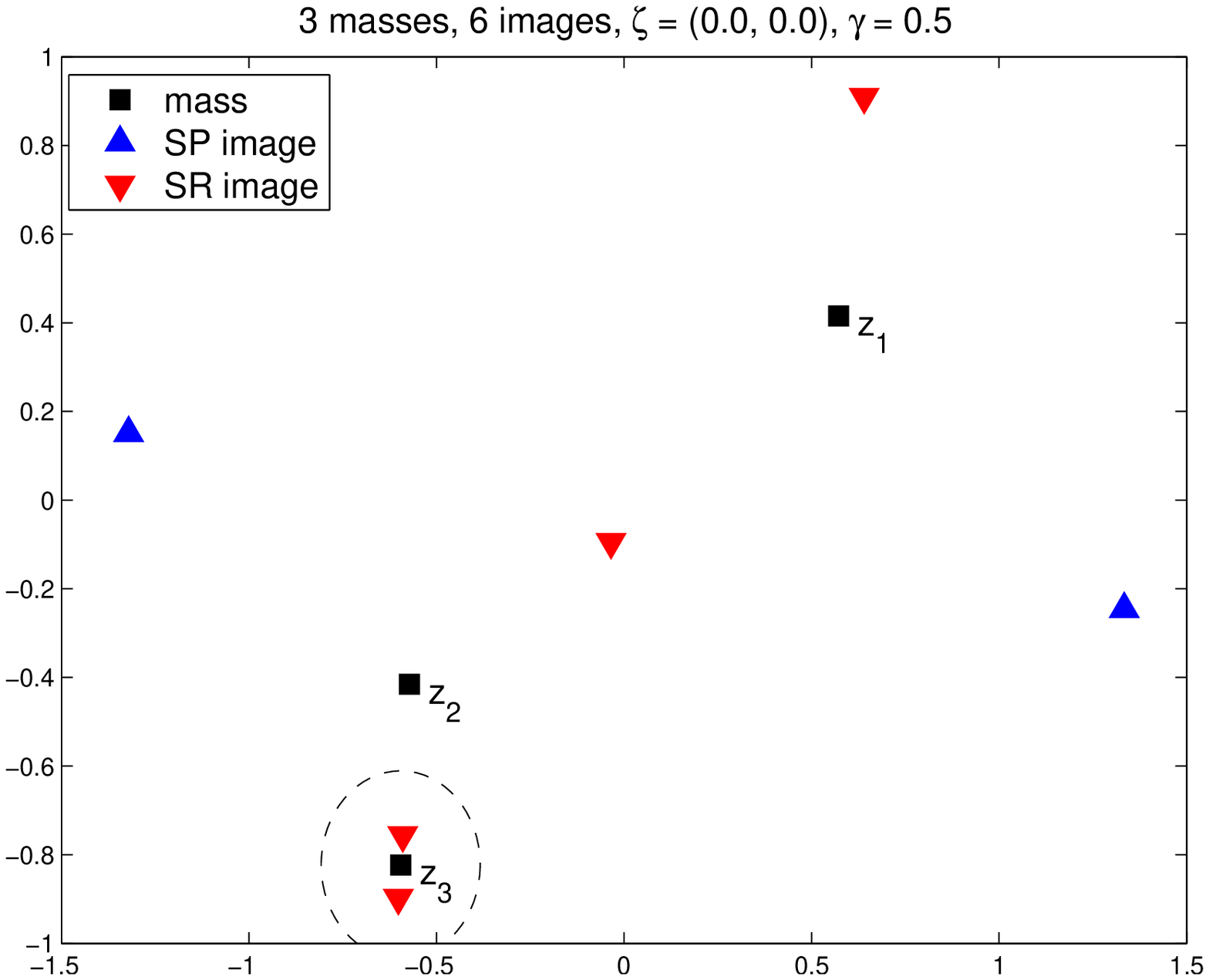}
\hfill
\includegraphics[width=.48\textwidth]{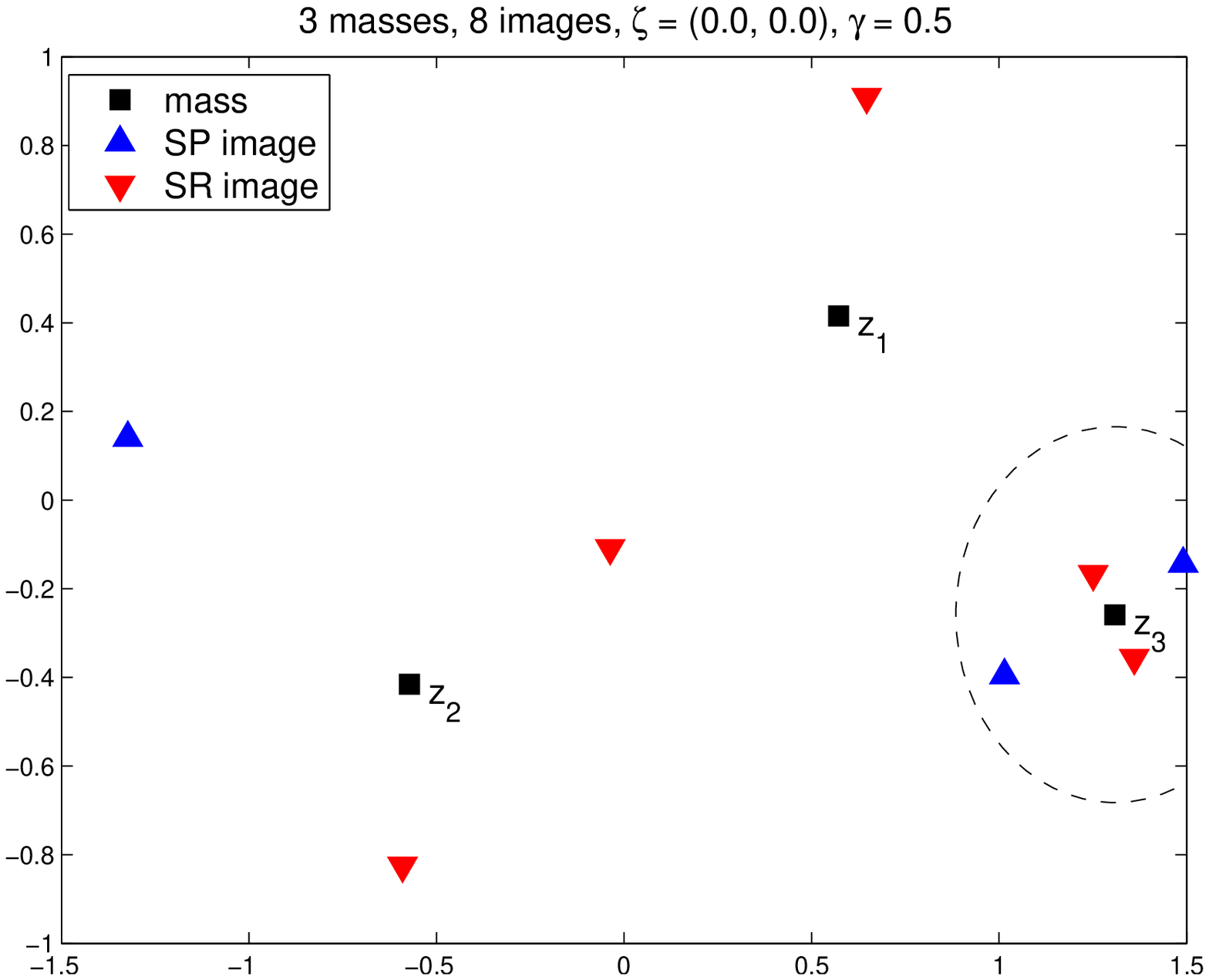}

\caption{\label{fig:pert} Illustration for Theorem~\ref{thm:pert}.
The black squares indicate mass points, and triangles show the
location of the induced images of the light source.  The images are
classified by ``SP'' (sense-preserving, blue, upward pointing)
and ``SR'' (sense-reversing, red, downward pointing).
The initial binary lens is shown in the top left picture.  The other
pictures show the lens after adding a mass at the indicated point
$z_3$, which is no image (top right), a sense-reversing image (bottom
left) and a sense-preserving image (bottom right) of the initial lens.
}
\end{figure}

\begin{rem}
\begin{compactenum}
\item The lens~\eqref{eqn:BD_lens} in the introduction is covered by
case~\ref{case:sp_img_zeroder}  in the preceding theorem with $m_{n+1}
= \eps$ and $d=n$.
For this lens, appropriate values of $m_{n+1}$ are
completely characterized in~\cite{LuceSeteLiesen2014a}.

\item In cases \ref{case:sr_img}--\ref{case:sp_img_zeroder}, the point 
$z_{n+1}$ is --of course-- no longer a solution of the lens equation 
$\eta_{n+1}(z) = \zeta$.

\item The lenses modeld by $\eta_n$ and $\eta_{n+1}$ have the same
number of images outside $D$, and these images are located at
approximately the same positions and retain their type
(sense-reversing or sense-preserving).

\item If one and two images are created in
cases~\ref{case:no_img} and~\ref{case:sr_img}, respectively, these images are
sense-reversing.  In cases~\ref{case:sp_img}
and~\ref{case:sp_img_zeroder} an equal number of sense-reversing and
sense-preserving images are created; see~\cite[thms.
3.1,3.14]{SeteLuceLiesen2014}.

\item In all cases of the above theorem, it is guaranteed that ``at
least'' a certain number of images are created (provided that $m_{n+1}$
is sufficiently small).  We believe that in fact no more than the stated number 
of images are created, and extensive numerical experiments support this
claim.  A mathematical proof of this claim is, however, a topic of
future research.

\item The effect of adding a mass larger than the ``sufficiently
small'' mass in the previous theorem
is twofold:  Either the mass is so
large that the lens is globally affected and images of $\eta_n$ ``far
away'' from $z_{n+1}$ may disappear.  Otherwise, even if the effect is
still local to $z_{n+1}$ with respect to $\eta_n$, \emph{more} than the
claimed number of images may be created; see~\cite[sect.
4.2]{SeteLuceLiesen2014} or~\cite[fig. 5]{LuceSeteLiesen2014a}.  This
effect is also shown in the example in
section~\ref{sec:maximal_lensing}.  A quantification of this effect
is subject of further research; see also~\cite{SeteLuceLiesen2014}.

\item The proofs in~\cite{SeteLuceLiesen2014} show that in each of the
cases the created images are located nearby (possibly rotated) roots
of unity with radius \emph{approximately} $\sqrt{m_{n+1}}$.  The smaller
the added mass $m_{n+1}$ is, the closer the images assume these
positions.  The radii of the new images can be quantified;
see~\cite[thm. 3.1]{SeteLuceLiesen2014}.
\end{compactenum}
\end{rem}

\medskip

Numerical examples for the cases \ref{case:no_img}--\ref{case:sp_img}
are shown in Fig.~\ref{fig:pert}.  The initial lens (top left image)
is a binary lens with masses $m_1 = 0.6$ and $m_2 = 0.4$, external
shear $\gamma = 0.5$, and a light source at $\zeta = 0$.  The other
three pictures (top right, bottom left, bottom right) display the
effect of adding a small mass of $m_3 = 0.02$ at a position $z_3$ for
each of the cases~\ref{case:no_img}--\ref{case:sp_img}.  As implied by
Theorem~\ref{thm:pert}, one, two and four images are created nearby
the mass position $z_3$, while the other images only alter their
positions slightly.
Examples for the case \ref{case:sp_img_zeroder} with exactly one
vanishing derivative of $R$ are given in
Section~\ref{sec:maximal_lensing}.

\section{Construction of maximal point lenses}
\label{sec:maximal_lensing}

In the introduction we noted that the only known examples for maximal
point lens models seem to arise from modifications of the point
lens of Mao, Petters and Witt~\cite{MaoPettersWitt1999}.
In this section we show how to construct from a given maximal point
lens on $n$ masses another maximal point lens on $n+1$ masses by
adding a tiny mass to the given lens.  The conditions we give are
in fact a special case of Theorem~\ref{thm:pert}, but applying this
theorem to this \emph{maximal lensing} case simplifies the conditions
imposed on the lensing map $\eta_n$ considerably and deserves a
statement on its own.  In the statement of this theorem we set the
external shear $\gamma$ to zero, but an analogous statement holds for 
gravitational lenses with external shear.

\begin{thm}
\label{thm:maxpert}
Let $\eta_n(z) = z - \sum_{j=1}^n \frac{m_j}{\conj{z} - \conj{z}_j} = z
- \conj{R(z)}$ model a point lens on $n$ masses $m_j>0$ at positions
$z_j \in \C$, that produces the maximal number of $5n - 5$ images.  Let $z_{n+1}
\in \C$ be an image with $R'(z_{n+1}) = 0$.   Then for all sufficiently small
masses $m_{n+1} > 0$, the lens modeled by the lensing map $\eta_{n+1}(z) =
\eta_n(z) - \frac{m_{n+1}}{\conj{z} - \conj{z}_{n+1}}$, which is of degree
$n+1$, produces the maximal number of $5n$ images, provided that the
source $\zeta$
does not lie on a caustic of $\eta_n$ or $\eta_{n+1}$.
\end{thm}

\begin{figure}[t]
\includegraphics[width=.48\textwidth]{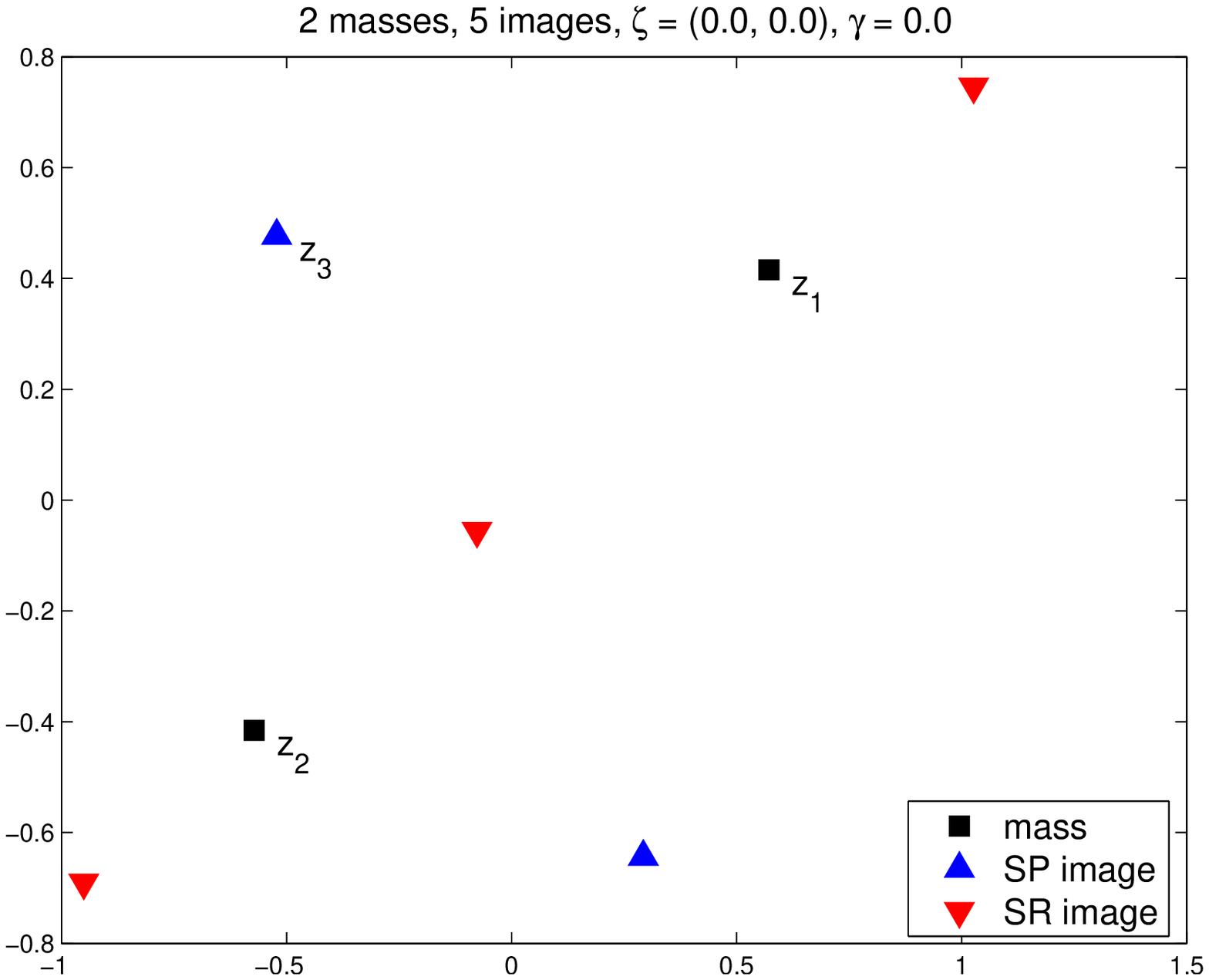}
\hfill
\includegraphics[width=.48\textwidth]{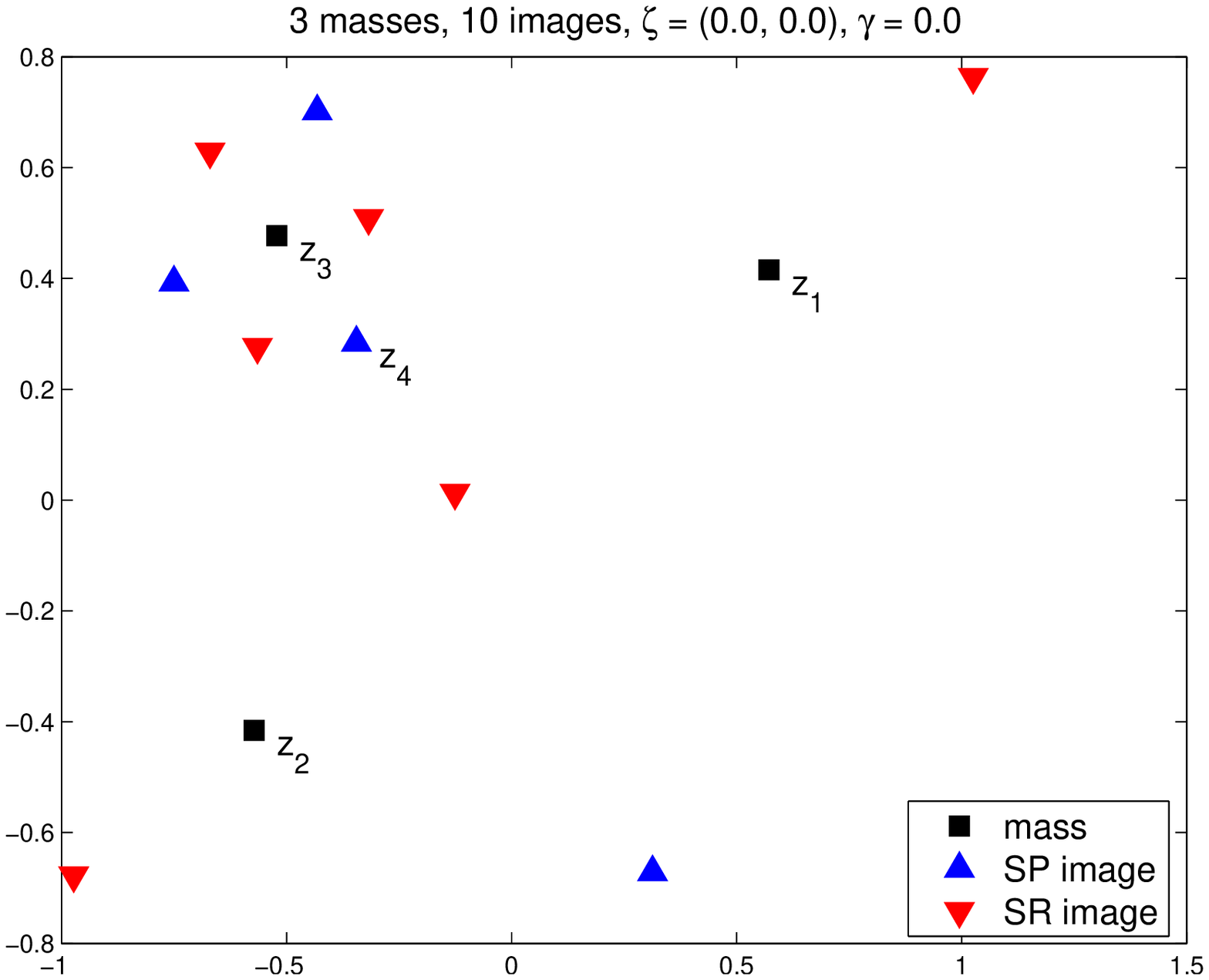}
\newline
\includegraphics[width=.48\textwidth]{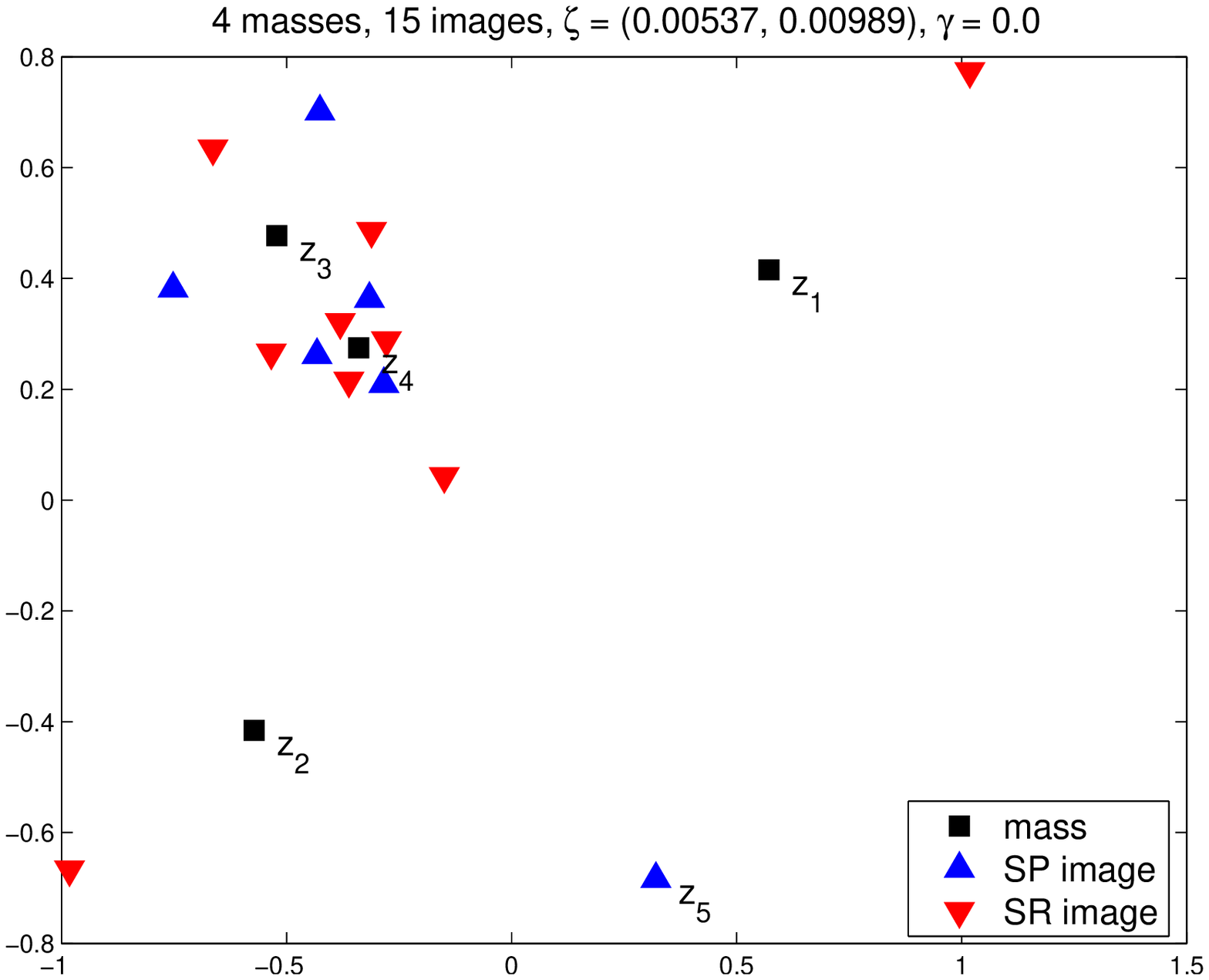}
\hfill
\includegraphics[width=.48\textwidth]{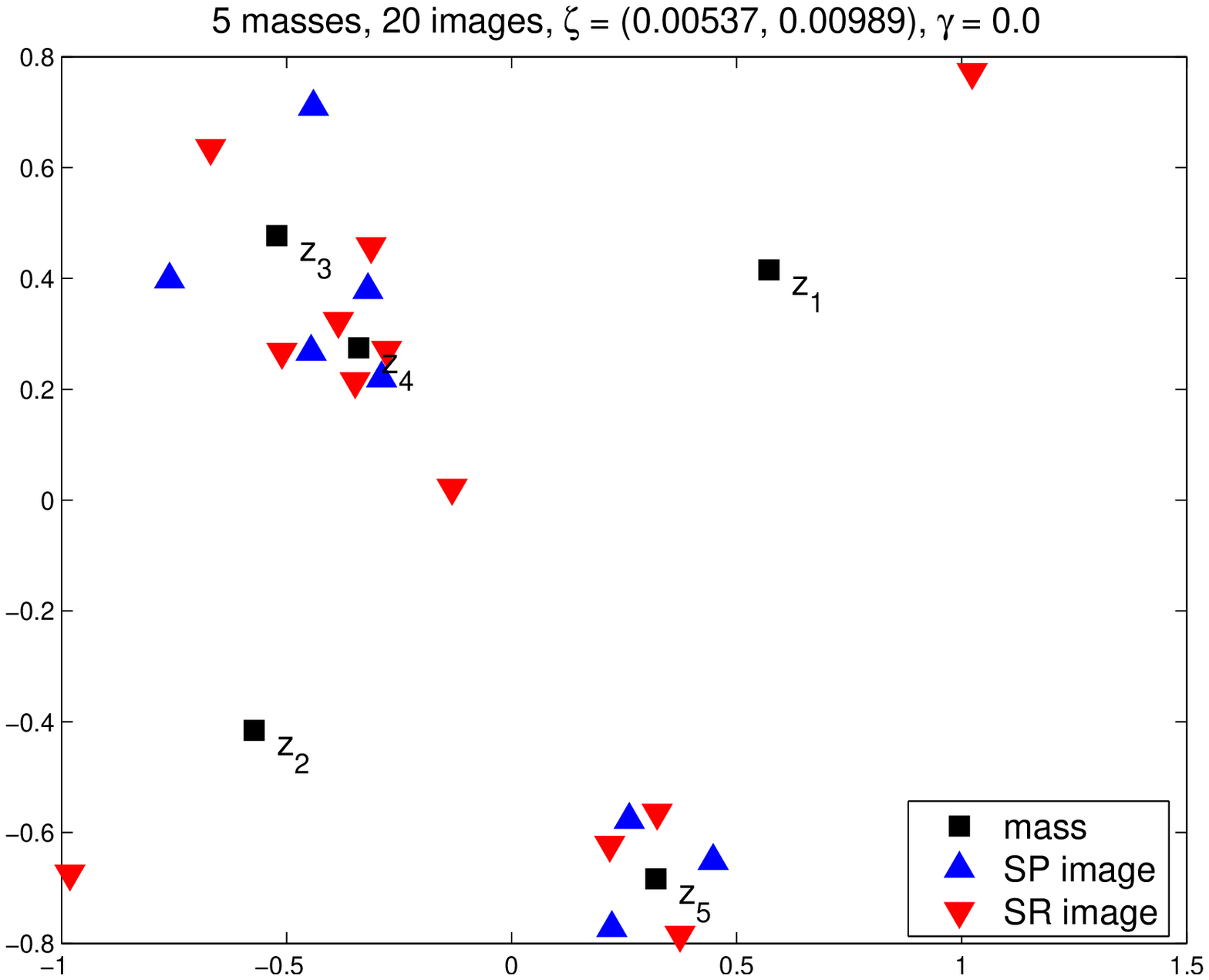}
\caption{Numerical example for the image-creating effect of adding
small masses at certain images of a maximal lens.  See
Section~\ref{sec:maximal_lensing} for a detailed discussion.  The
symbols used are the same as in Fig.~\ref{fig:pert}.
\label{fig:maximal_lensing}}
\end{figure}

Note that $R''(z_{n+1}) \neq 0$, because $\eta_n$ is a maximal lens.
Also, the image $z_{n+1}$ is
necessarily an unmagnified image, since
the magnification of $z_{n+1}$ is
\begin{equation*}
\mbox{Mag}(z_{n+1}, \zeta)\,
    = \abs{\det \jac \eta(z_{n+1})}^{-1}
    = \abs{1 - \abs{R'(z_{n+1})}^2}^{-1}
    = 1.
\end{equation*}

The theorem is illustrated in Figure~\ref{fig:maximal_lensing}.  The
initial lens is depicted in the top left image.  This binary lens with
masses $m_1 = 0.6$ and $m_2 = 0.4$ at positions $z_1$ and $z_2$ and
zero external shear produces five images of the source, thus it is a
maximal lens.  The projected position of the light source $\zeta$ on
the lens plane is the origin.  At the image $z_3$ indicated in the
plot, we have $\abs{R'(z_3)} \approx 5 \cdot 10^{-16}$, so the image
$z_3$ satisfies (numerically) the condition of
Theorem~\ref{thm:maxpert}.

The result of adding a third mass of $m_3 = 0.05$ at $z_3$ is shown
in the top right picture.  As implied by the theorem, six ``new''
images around the newly created mass appear.  As the ``old'' images
only alter their positions slightly, but no image disappears (except
for $z_3$), the
constructed lens is maximal again.

After adding the mass $m_3$ at the position $z_3$ to the lens plane,
the derivative of $R$ does not vanish at any of the ten images.
However, for the image $z_4$ indicated in the plot, we have
$\abs{R'(z_4)} \approx 0.0954$, which is already quite small.  By
shifting the projected source position $\zeta$ slightly within the
caustic from $(0,0)$ to
approximately $(0.00537,0.00989)$, the image $z_4$ moves to a nearby
point in the lens plane, at which $R'$ vanishes.  Adding a mass of
size $m_4 = 0.005$ at this displaced image again produces $6$
new images in the vicinity of the newly added mass, and thus we have
constructed a maximal point lens on four masses.  The resulting lens
configuration is shown in the bottom left plot.

Finally we wish to emphasize that the condition specified in
Theorem~\ref{thm:maxpert}, viz., that the derivative must vanish at
the point where the mass is to be added, is only \emph{sufficient} for
obtaining again a maximal lens, but not necessary.  It may well be
that the maximum number of images, six, is already produced if the
derivative is sufficiently small.  This aspect is exemplified by the
image $z_5$ in the bottom left plot.  Here we have $\abs{R'(z_5)}
\approx 0.09$, but yet adding a small mass of $m_5 = 0.015$ produces
six new images.  Fewer images (four) are created, however, if $m_5$ is
somewhat smaller, as implied by case~\ref{case:sp_img} of
Theorem~\ref{thm:pert}.  The resulting maximal lens is shown in the
bottom right picture.

\section{Conclusions and outlook}

In this note we have presented a complete characterization of the
image creating effect when a mass is inserted into a given
microlensing model.  The assumptions in the mathematical assertions
cover microlensing models with and without shear and are applicable to
any number of point masses.  Our findings generalize a particular
construction for maximal point lenses by Rhie and we have given
a general methodology for the construction of maximal point lens
models.

The two most important open questions in the context of
Theorem~\ref{thm:pert} are, firstly, to prove that the lower bounds on
the number of created images are in fact equalities and, secondly,
to quantify the allowable mass such that the claimed number of images
are created.

Finally we mention that the question of maximal lensing in models with
objects of radial mass density is much less understood than in point
lens models~\cite{KhavinsonLundberg2011}.

\paragraph*{Ackowledgements}
Robert Luce's work is supported by Deutsche Forschungsgemeinschaft,
Cluster of Excellence ``UniCat''.

\bibliographystyle{plain}
\bibliography{ulens_pert}

\end{document}